# Using Partial Structure R1 to Determine Small-Molecule Crystal Structures at the Lowest Resolution Limits

Authors

**Xiaodong Zhang[a]**

[a]Chemistry Department, Tulane University, 6400 Freret Street, New Orleans, Louisiana, 70118, United States

Correspondence email: xzhang2@tulane.edu

**Synopsis**  The lowest data resolution at which the partial structure R1 (pR1) method can determine a small-molecule crystal structure has been studied for three specific examples.

**Abstract**  This paper explores the lowest data resolution at which the partial structure R1 (pR1) method can determine a small-molecule crystal structure. Three specific structures have been studied. For a structure having 4 S atoms out of total 32 atoms in its unit cell, the lowest structure-solvable resolution is 1.5 Å, and at this resolution all the atoms are individually searchable by the sR1 method (a specialized form of the pR1 method in which the partial structure contains a single atom). For the other two samples, which have 46 and 156 C atoms in their unit cells, respectively, the lowest structure-solvable resolutions are 1.5 Å and 1.2 Å, respectively, and both require a combination of the pR1 and the sR1 methods to determine their structures at these resolution limits. In all three cases, the usual choice of method for solving a small-molecule crystal structure, namely, the *SHELXT* program, is not applicable. Therefore, the pR1 method complements the *SHELXT* for determination of a small-molecule crystal structure at these lowest resolution limits.

## 1. Introduction

A recent paper (Zhang & Donahue, 2024) has introduced the single-atom R1 (sR1) method for solving small-molecule crystal structures, as well as the more general concept, the partial structure R1 (pR1). The pR1 method (or the special form, the sR1 method, in which the partial structure contains a single atom) is a new model-searching technique that is used to directly solve a small-molecule crystal structure. In comparison, the traditional model-searching techniques only serve the initial phasing purpose in the molecular-replacement (MR) method (Rossmann & Blow, 1962; Crowther, 1972;

Rossmann, 1972, 1990; Bricogne, 1992; Read, 2001; McCoy, 2004; McCoy *et al.*, 2017; Caliandro *et al.,* 2009; Burla, *et al.*, 2020), and the MR is typically applied in the macromolecular crystallography. Recently, Gorelik *et al.* (2023) have evaluated the MR as implemented in *phaser* (McCoy *et al.*, 2007) for small-molecule crystal structure determination from X-ray and electron diffraction data with reduced resolution. Their work has inspired us to ask this question: what is the lowest data resolution at which the pR1 method can determine a small-molecule crystal structure? In this report we use the same examples (see Table 1) as previously studied (Zhang & Donahue, 2024) to shed some light on the answer to this question. The result of this study indicates that the pR1 method complements the usual choice of program for solving small-molecule crystal structures, namely, the *SHELXT* program (Sheldrick, 2015), when the data resolution is very low.

**Table 1**  Crystallographic information of the samples as previously studied (Zhang & Donahue, 2024)

| Sample | Formula (excluding H) | Z | non-hydrogen atoms in cell | a(Å) | b(Å) | c(Å) | α(°) | β(°) | γ(°) | space group |
|---|---|---|---|---|---|---|---|---|---|---|
| 1 | $S_2O_2C_{12}$ | 2 | 32 | 5.86 | 10.34 | 10.74 | 90 | 104.50 | 90 | P2(1) |
| 2 | $C_{46}$ | 1 | 46 | 5.95 | 10.80 | 12.97 | 103.77 | 99.95 | 90.46 | P-1 |
| 3 | $C_{78}$ | 2 | 156 | 12.34 | 15.98 | 16.57 | 114.10 | 90.70 | 103.20 | P-1 |

**2. Use the sharpened reflection data in the pR1 and/or the sR1 calculation**

The same implementation (Zhang & Donahue, 2024) of the pR1 and the sR1 methods is used in this report, except for one improvement: instead of directly feeding the raw reflection intensity $F_o^2(hkl)$ to the calculation, the sharpened intensities are fed. A sharpened intensity is calculated as a product of multiplying the raw intensity $F_o^2(hkl)$ with $\exp(2Bs^2)$, where $s=\sin\theta/\lambda$. The parameter B is determined by the Wilson method (see our implementation of the Wilson method in the supporting information). We have found that if the sharpened intensities are used, there are fewer ghost atoms during an sR1 calculation. This is understandable, because the sR1 (as well as the pR1) method employs an atomic model with no thermal effect (i.e., using the isotropic displacement parameter U=0 Å$^2$), and such a model matches the sharpened data better (because sharpening removes the average thermal effect).

**3. The lowest resolution at which the sR1 method can determine the structure of sample 1**

The raw data resolution of sample 1 is 10.4 – 0.84 Å. We have found that when the data is clipped at a resolution as low as 1.5 Å, the sR1 method can determine the structure of sample 1 (see details in the supporting information). This means all the single atoms of sample 1 are searchable one at a time even at 1.5 Å resolution. Sample 1 contains four S atoms in its unit cell. These heavy atoms must have contributed greatly to the success of the sR1 method in this case. Heavy atoms are easier to be located, and once they have been located, the further sR1 calculation becomes more reliable.

**4. The lowest resolution at which the combination of the pR1 and the sR1 methods can determine the structure of sample 2**

Sample 2 has 46 C atoms in its unit cell. For samples like 2 and 3, which have no heavy atoms, at a low resolution, the sR1 method alone cannot determine the structure. Previous study (Zhang & Donahue, 2024) indicates that in such difficult cases one may use the pR1 to put a known fragment to jumpstart the sR1 calculation. To reach the lowest resolution limit, the fragment used to jumpstart the calculation should be as large as possible. The largest known fragment that samples 2 and 3 contain is shown in figure 1, which consists of a central benzene ring with 6 C atoms bonding to the 6 corners. We call this fragment as a benzene-star. All the C-C bonds in a benzene-star have bond length 1.39 Å.

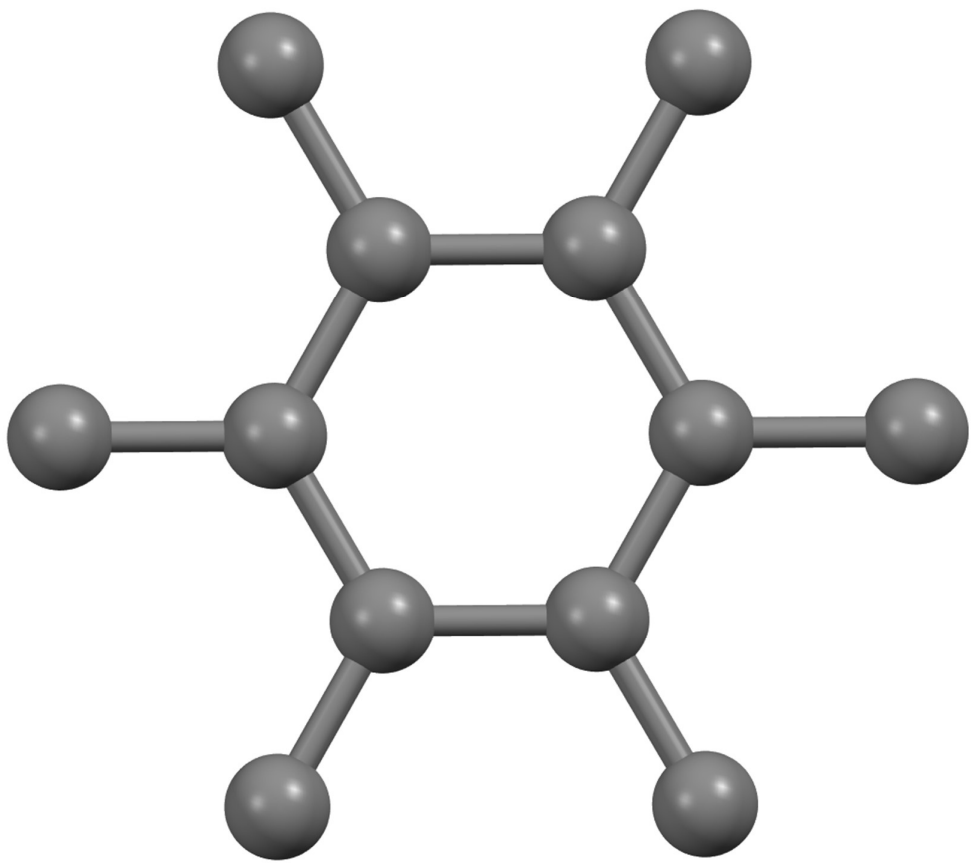

**Figure 1** The structure of a benzene-star fragment. All the C-C bonds in this fragment have bond length 1.39 Å.

The pR1 method can determine the correct orientation(s) of a known fragment (Zhang & Donahue, 2024). It can also position a fragment of a known orientation by globally minimizing the pR1 while adjusting the (x,y,z) of the local origin of the fragment. The technical details of these calculations are given in the supporting information.

The raw data resolution of sample 2 is 12.39 – 0.73 Å. We have found that the lowest data resolution at which this structure can be determined is 1.5 Å, and the method for determining this structure at this resolution limit is: first put a benzene-star that is correctly oriented by the pR1 method at a position where its center is at (0.3,0.3,0.3), and then complete the structure with the sR1 calculations (more details of these steps are provided in the supporting information). The determined structure, when compared to the correct structure, has all 46 atoms located within 0.5 Å. So, the combination of the pR1 and the sR1 methods has successfully determined the structure of sample 2 at a data resolution of 1.5 Å.

**5. The lowest resolution at which the combination of the pR1 and the sR1 methods can determine the structure of sample 3**

The raw data resolution of sample 3 is 15.02 – 0.77 Å. We have found that the lowest data resolution at which this structure can be determined is 1.2 Å. In this case, though it is also possible to jumpstart the sR1 calculation with a single benzene-star, the calculation is extremely difficult to complete. The calculation becomes much easier if it is initiated with two, or three, or four benzene-stars. Figure 2 shows the case of using four benzene-stars to initiate the calculation. There are two steps. In the first step, we add four benzene-star fragments. These benzene-stars share two orientations (each orientation is shared by a pair of benzene-stars). These orientations are determined by using the pR1 of a free-standing benzene-star (free-standing means a model containing a single fragment with no other known atoms). Technical details of this type of calculation are available in the supporting information. Once these orientations (labeled as 0 and 1) are determined, we position the four fragments one at a time. The first benzene-star has orientation 0 and can be positioned with its center at (0.3,0.3,0.3). The second and the third benzene-stars have orientation 1 and are each positioned by globally minimizing the pR1 while adjusting the (x,y,z) of the center of the benzene-star. The fourth benzene-star has orientation 0 and is also positioned by globally minimizing the pR1. In the second step, we complete the model by locating the other missing C atoms one at a time by globally minimizing the sR1 while adjusting the (x,y,z) of a missing C atom. The second step includes multiple cycles of manually deleting the ghost atoms and re-extending to the full model with the sR1 calculations. The final model has only two C atoms being misplaced. Thus, the combination of the pR1 and the sR1 methods has successfully determined the structure of sample 3 at a data resolution of 1.2 Å.

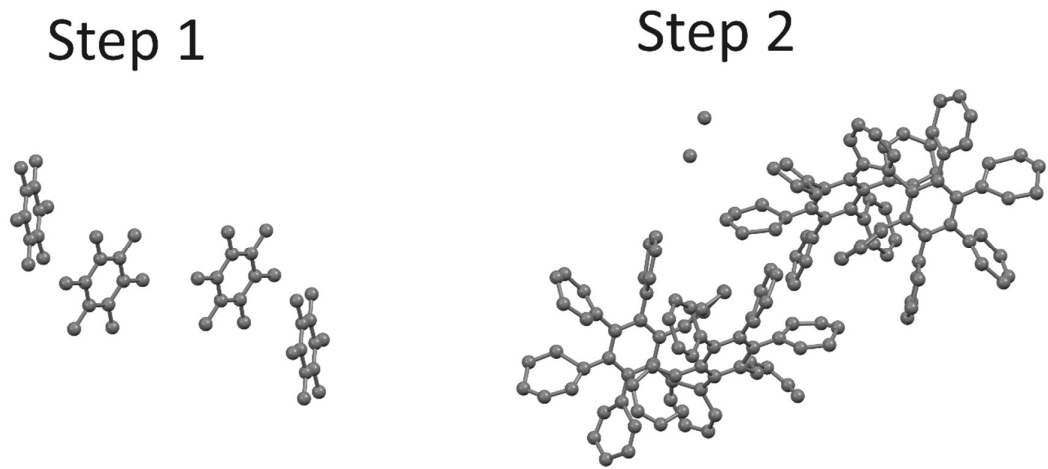

**Figure 2** The steps of determining the structure of sample 3 when data is clipped at the resolution of 1.2 Å. Step 1: use the pR1 method to correctly orient and position 4 benzene-star fragments. Step 2: use the sR1 method to complete the model.

## 6. Conclusions and Discussions

We have demonstrated that, for a sample having 4 S atoms (sample 1), at the lowest data resolution of 1.5 Å, all its 32 atoms in its unit cell can be searched one at a time by the sR1 method. For the other two samples (2 and 3), which have only C atoms, at their lowest resolution, directly searching single C atoms by the sR1 method is unsuccessful (this has not been shown in this report). For sample 2, which has 46 C atoms in its unit cell, at its lowest resolution of 1.5 Å, a successful single atom search can be jumpstarted by putting a single benzene-star that has been correctly oriented by the pR1 method. For sample 3, which has 156 C atoms in its unit cell, at its lowest resolution of 1.2 Å, though it is possible to jumpstart the single-atom search with a single benzene-star, the calculation is much easier if it is initiated by two or three or four benzene-stars which are correctly oriented and positioned by the pR1 method. All these observations lead to a conclusion that the presence of heavy atoms and/or large known fragment(s) is critical to the success of the pR1 method for determination of a small-molecule structure when the data resolution is very low.

The usual choice of method for solving small-molecule crystal structures is the *SHELXT* program (Sheldrick, 2015). Though this program can successfully solve the structures of samples 1, 2, and 3 at their raw data resolutions or at resolutions slightly lower, it does not yield meaningful results when the data are clipped at the lowest resolution where the sR1 method or the combination of the pR1 and the sR1 methods can still work. Details of using *SHELXT* to solve these structures are available in the supporting information. Therefore, the pR1 (which includes the sR1 as a special form) method complements the *SHELXT* program at these lowest-resolution regions. No doubt, the pR1 method is a viable choice for tackling the situations where a data resolution could be low, for example in the electron diffraction experiments (Gorelik *et al.*, 2023). (Though our work has been inspired by Gorelik *et al.*'s work, unfortunately, as explained in the supporting information, a proper comparison between our work and their work cannot be made at this point.)

**Acknowledgements**   Professor Robert A. Pascal has kindly provided the data sets of samples 2 and 3.

**References**

Bricogne, G. (1992). *Proceedings of the CCP4 Study Weekend. Molecular Replacement*, edited by W. Wolf, E. J. Dodson & S. Gover, pp. 62-75. Warrington: Daresbury Laboratory.

Burla, M. C., Carrozzini, B., Cascarano, G. L., Giacovazzo, C. & Polidori, G. (2020). *Acta Cryst.* D**76**, 9-18.

Caliandro, R., Carrozzini, B., Cascarano, G. L., Giacovazzo, C., Mazzone, A. & Siliqi, D. (2009). *Acta Cryst.* A**65**, 512-527.

Crowther, R. A. (1972). In *The Molecular Replacement Method*, Ed. M. G. Rossmann, Gorden & Breach, New York, pp. 173-178.

Gorelik, T. E., Lukat, P., Kleeberg, C., Blankenfeldt, W. & Mueller, R. (2023). *Acta Cryst*. A**79**, 504-514.

McCoy, A. J. (2004). *Acta Cryst.* D**60**, 2169-2183.

McCoy, A. J., Grosse-Kunstleve, R. W., Adams, P. D., Winn, M. D., Storoni, L. C. & Read, R. J. (2007). *J. Appl. Cryst*. **40**, 658–674.

McCoy, A.J., Oeffner, R. D., Wrobel, A. G., Ojala, J. R., Tryggvason, K., Lohkamp, B. & Read, R. J. (2017). Ab initio solution of macromolecular crystal structures without direct methods. *Proc. Natl. Acad. Sci. USA* **114**, 3637-3641.

Read, R. J. (2001). *Acta Cryst.* D**57**, 1373-1382.

Rossmann, M. G. (1972). Editor. *The Molecular Replacement Method*. New York: Gordon & Breach.

Rossmann, M. G. (1990). *Acta Cryst.* A**46**, 73-82.

Rossmann, M. G. & Blow, D. M. (1962). *Acta Cryst*. **15**, 24-31.

Sheldrick, G. M. (2015). *Acta Cryst.* A**71**, 3-8.

Zhang, X. & Donahue, J. P. (2024). https://doi.org/10.1107/S2053273324001554.

# Supporting information

### S1. Information on data collection of the samples

All crystals were coated with paratone oil and mounted on the end of a nylon loop attached to the end of the goniometer. Data were collected at 150 K under a dry $N_2$ stream supplied under the control of an Oxford Cryostream 800 attachment. The data collection instrument was a Bruker D8 Quest Photon 3 diffractometer equipped with a Mo fine-focus sealed tube providing radiation at $\lambda = 0.71073$ nm or a Bruker D8 Venture diffractometer operating with a Photon 100 CMOS detector and a Cu Incoatec I microfocus source generating X-rays at $\lambda = 1.54178$ nm.

### S2. Our implementation of the Wilson method for determination of the parameter B

The raw reflection intensities $F_o^2(hkl)$ are ranked by $s=\sin\theta/\lambda$ and are divided into either 20 groups if total number of reflections exceeds 10000 or 10 groups otherwise. For each group, we calculate the average intensity which we denote as $<F_o^2>$, and we also calculate the average s value which we denote as $<s>$. This leads to the following approximate equation:

$$k <F_o^2> \approx [\sum_j f_j^2(<s>)]\exp(-2B<s>^2)$$

In this equation, k is a scaling factor for the observed intensities, and B is the parameter whose value we are seeking for.

This equation can convert into a linear form of Y=a+bX, in which X and Y are defined as:

$$X = <s>^2$$

$$Y = ln\frac{<F_o^2>}{\sum_j f_j^2(<s>)}$$

and the intercept a and the slope b are:

$$a = -\ln(k)$$

$$b = -2B$$

After calculating the (X,Y) data points for all the groups we use linear fit to find the intercept a and the slope b, and further calculate the parameters k and B as follows:

$$k = \exp(-a)$$

$$B = -b/2$$

Once the parameter B is obtained, a sharpened intensity is calculated as:

$$(F_o^2)_{sharpene} = F_o^2 \exp(2Bs^2)$$

However, the sharpened intensities are not scaled by multiplying with the scaling factor k; instead, they are scaled such that their sum equals $\Sigma_{hkl} \Sigma_j f_j^2(hkl)$.

### S3. The lowest resolution limit for sample 1

The sR1 method is used to solve the structure of sample 1. The calculation strategy is: starting from a single S atom at (0.3,0.3,0.3), use the sR1 to expand to 4 atoms, to 12 atoms, to 24 atoms, and to 32 atoms. There are a few cycles of deleting the ghost atoms and re-extending to the full model.

We have found that if the data resolution is clipped at 1.6 Å the structure cannot be solved, but if clipped at 1.5 Å the structure can be solved to give the following model:

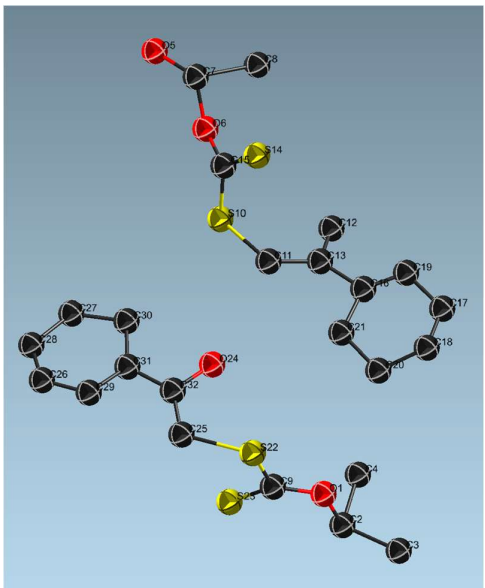

Compared with the correct model, this resulting model has all 32 atoms being located within 0.5 Å. (The algorithm for making this type of comparison is given in the next section.)

### S4. The algorithm of comparing two models

This algorithm only compares the positions of the atoms in the two models; it disregards the types of atoms. Let a, b, and c be the length of the unit cell edges. One difficulty of making this comparison is that the two models may have different locations relative to the unit cell edges. The algorithm uses an approximate method to overcome this difficulty: shift all atoms of one model (model A) such that one of its atoms overlaps one atom of the other model (model B). Try this for all atoms in both models. So, the calculation needs to be repeated for $N_A \times N_B$ times, where $N_A$ and $N_B$ are the number of atoms in the models A and B, respectively. Further these calculations also need to be repeated after inverting model B, in case model A matches the inverted version of B better. After repeating the calculation, use the result of the best match that has been discovered.

After overlapping one particular atom of model A with a particular atom of model B by properly shifting all atoms of model A, we make sure all atoms of both models are located within the unit cell (by adding 1 to or subtracting 1 from the fractional coordinates x,y,z of an atom). With such preparation we are ready to count how many atoms of A are each within s = 0.5 Å of an atom in B. To make such comparison quickly, we convert the fractional coordinates x,y,z of an atom to integers in following way:

$N_x$ = int(a/s), $N_y$ = int(b/s), $N_z$ = int(c/s)

$I_x$ = int($xN_x$), $I_y$ = int($yN_y$), $I_z$ = int($zN_z$)

Thus, ($I_x$, $I_y$, $I_z$) is the integer coordinate of an atom.

Prepare a mask $M_A(i, j, k)$ for model A, where i=0 to $N_x$-1, j=0 to $N_y$-1, and k=0 to $N_z$-1. $M_A(i, j, k)$ takes value 0 unless (i,j,k) is the integer coordinate of an atom. Similarly, $M_B(i, j, k)$ is a mask for model B.

The number of atoms of model A which overlaps within 0.5 Å with an atom of B can be calculated as $\sum_{i,j,k} M_A(i,j,k) M_B(i,j,k)$.

## S5. Technical details of using pR1 to determine the orientation of a fragment as well as to position a fragment of a known orientation

A local Cartesian coordinate system is set up for the fragment by selecting a local origin and three orthogonal directions.

A Cartesian coordinate system is also set up for the unit cell. The origin of this system overlaps the origin of the unit cell. Its three unit vectors **x**, **y**, and **z** are related to the cell vectors **a** and **b** as follows: **x** = **a**/|**a**|, **y** = (**b**-**b**·**xx**)/| **b**-**b**·**xx** |, **z** = **x**×**y**. The Cartesian coordinates and the fractional coordinates of an atom in the unit cell are inter-converted during the calculations. Using cell parameters $a$, $b$, $c$, $\alpha$, $\beta$, and $\gamma$, the relation between fractional coordinates ($x_f, y_f, z_f$) and Cartesian coordinates ($x_c, y_c, z_c$) is expressed as:

$$\begin{pmatrix} x_c \\ y_c \\ z_c \end{pmatrix} = \begin{pmatrix} a & b\cos\gamma & c\cos\beta \\ 0 & b\sin\gamma & \frac{c(\cos\alpha - \cos\beta\cos\gamma)}{\sin\gamma} \\ 0 & 0 & \frac{V}{ab\sin\gamma} \end{pmatrix} \begin{pmatrix} x_f \\ y_f \\ z_f \end{pmatrix}$$

in which the cell volume V is calculated as:

$$V = abc\sqrt{1 - \cos^2\alpha - \cos^2\beta - \cos^2\gamma + 2\cos\alpha \times \cos\beta \times \cos\gamma}$$

At start, a fragment is positioned such that its local Cartesian system overlaps the Cartesian system of the unit cell. The fragment is attached to its local Cartesian system. So, rotation and translation of the fragment is realized by rotating and translating its local Cartesian system in the cell Cartesian system. Translation is realized by translating the local origin in the cell Cartesian system (or rather in the cell fractional coordinate system, and the fractional coordinates and the cell Cartesian coordinates are inter-converted). A general rotation is consisted by three simple rotations: a rotation around x-axis (of the local frame) in the direction from y-axis to z-axis through angle $\psi$; a rotation around z-axis (of the local frame) in the direction from x-axis to y-axis through angle $\varphi$; and another rotation around x-axis (of the local frame) in the direction from y-axis to z-axis through angle $\zeta$. Before rotation, a point has Cartesian coordinates $(x,y,z)$. After a rotation of angles $(\psi,\varphi,\zeta)$, the point moves to a new location in the same Cartesian system with Cartesian coordinates $(x',y',z')$, which are calculated by:

$$\begin{pmatrix} x' \\ y' \\ z' \end{pmatrix} = \begin{pmatrix} 1 & 0 & 0 \\ 0 & \cos\psi & -\sin\psi \\ 0 & \sin\psi & \cos\psi \end{pmatrix} \begin{pmatrix} \cos\varphi & -\sin\varphi & 0 \\ \sin\varphi & \cos\varphi & 0 \\ 0 & 0 & 1 \end{pmatrix} \begin{pmatrix} 1 & 0 & 0 \\ 0 & \cos\zeta & -\sin\zeta \\ 0 & \sin\zeta & \cos\zeta \end{pmatrix} \begin{pmatrix} x \\ y \\ z \end{pmatrix}$$

In general, to cover all possible rotations, the range of $\psi$ should be 0 to 360 degrees, the range of $\varphi$ should be 0 to 180 degrees, and the range of $\zeta$ should be 0 to 360 degrees. When a fragment has n-fold rotation symmetry and the rotation axis is along the x-axis of its local Cartesian system, then the range of $\zeta$ can reduce to 0 to 360/n degrees.

The space spanned by rotation and translation is a 6-dimensional orientation-location space. (It is 5-dimensional if the fragment is linear.) To coarsely locate the global minimum point or the local minimum points (the holes) of a pR1 map in this 6-dimensional space, a grid is set up with 0.4 Å step size in translations within the cell and 5-degree step size in rotation angles. The range of the rotation angles are: 0 to 360 degrees for $\psi$, 0 to 180 degrees for $\varphi$, and 0 to 360 degrees for $\zeta$. If the fragment has n-fold rotation symmetry, and the rotation axis is arranged along the local x-axis, the range of $\zeta$ can reduce to 0 to 360/n. The precision of locating the global minimum point or the local minimum points is refined by halving the step size locally five times.

The general hypothesis is that the deepest hole of a pR1 map in a 6-dimensional orientation-location space determines the orientation and location of a missing fragment (Zhang & Donahue, 2024). Note that by "location of a fragment" we mean the location of the local origin of the fragment (while keeping its orientation unchanged); similarly, "locating a fragment" means locating the local origin of a fragment. Determining pR1 holes in a 6-dimensional space is time-consuming. To cut calculation time, the problem is divided into two 3-dimensional calculations, that is, finding the orientation and the location in two separate steps.

For a free-standing fragment, that is, a model consisting of a single fragment with no other known atoms, the pR1 only depends on the orientation of the fragment. Therefore, the possible orientations of all missing fragments are detected by the holes in this pR1 map of a free-standing fragment in a 3-dimensional orientation space. Due to the possible symmetry of a fragment and the redundancy in representing an orientation with the three rotation angles, many orientation representations are equivalent to each other. Considering two orientation representations, if the atoms of the fragment of one orientation representation can match the atoms of the fragment of the other representation in a one-to-one basis within 0.25 Å (see the algorithm for this type of comparison in section S4), the two representations are considered equivalent. The non-equivalent orientation representations are filtered out from all detected representations. These serve as the candidate orientations.

The candidate orientations are ranked from the smallest R1 to the highest R1 and are labelled by 0, 1, 2, etc. In some situations, for example, if it is known that there are only two fragments in the structure, then it is obvious that one fragment has orientation 0 and the other has orientation 1. In such a situation, fragment 0 takes orientation 0, and its location is determined by the deepest hole in a pR1 map of a 3-dimensional location space. After determination of fragment 0, a new pR1 is defined by including the atoms of fragment 0 as known atoms and taking orientation 1 the location of fragment 1 is determined by the deepest hole of this new pR1 map in a 3-dimensional location space. In other situations, there are multiple fragments in the structure. In such situations, to determine a missing fragment, it is necessary to try all possible orientations. For each trial orientation, the best choice of location of a missing fragment is determined by the deepest hole in a pR1 map of a 3-dimensional location space. This best choice is combined with the trial orientation to form a candidate orientation-location. Trying all candidate orientations leads to a list of candidates of orientation-locations which can be ranked from the smallest R1 to the highest R1. The missing fragment is determined by the candidate orientation-location of the lowest R1. With this missing fragment being determined, a newer pR1 is defined by including the atoms of the newly determined fragment as known atoms. This newer pR1 is used to determine the next missing fragment in the same way.

When using the single atom R1 (sR1) to locate single missing atoms (Zhang & Donahue, 2024), a rule excluding clustering ghost atoms and a rule excluding triangular bonding are enforced. Similarly, here, when using pR1 to locate a missing fragment, these rules are also enforced: if a trial orientation-location for a missing fragment will cause one of its atoms being a clustering ghost atom, or one of its atoms will involve triangular bonding with two known atoms, this trial is disqualified as a candidate for determining the orientation-location of the missing fragment.

**S6. The lowest resolution limit for sample 2**

The calculation strategy for sample 2 is: first we determine the correct orientation of a benzene-star fragment by globally minimizing the pR1 of a free-standing benzene-star and position it such that its center is at (0.3,0.3,0.3), then use the sR1 method to complete the model.

We have found that if the data resolution is clipped at 1.6 Å the structure cannot be solved, but if clipped at 1.5 Å the structure can be solved to yield a completely correct model:

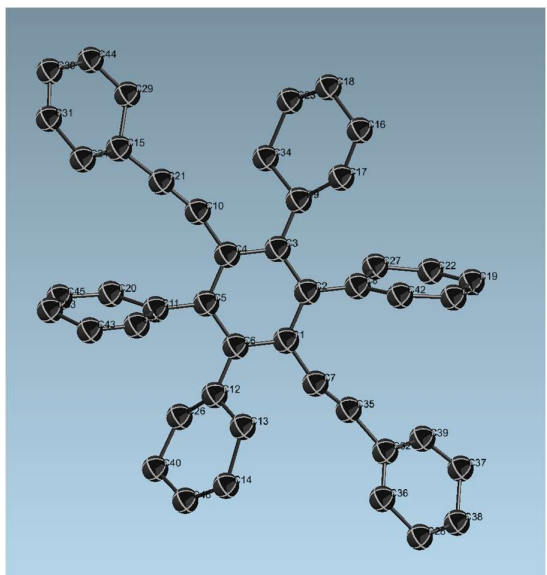

## S7. Use *SHELXT* program to solve the structures of samples 1, 2, and 3

The *SHELXT* program can correctly solve the structure of sample 1 if the raw data is used:

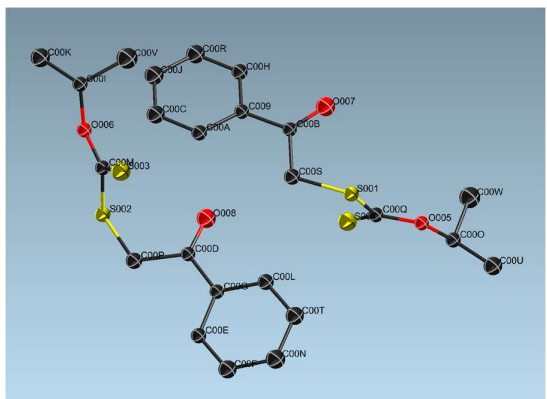

However, if the data is clipped at the resolution of 1.5 Å the program yields a meaningless result:

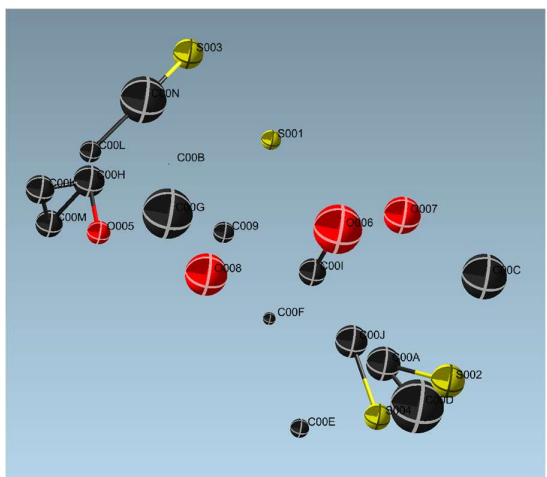

The *SHELXT* program can correctly solve the structure of sample 2 if the raw data is used:

However, if the data is clipped at the resolution of 1.5 Å the program yields a meaningless result:

The *SHELXT* program can correctly solve the structure of sample 3 if the raw data is used:

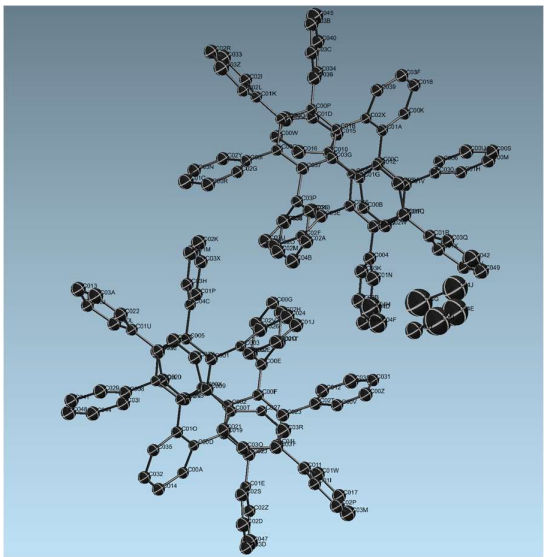

However, if the data is clipped at the resolution of 1.2 Å the program yields a meaningless result:

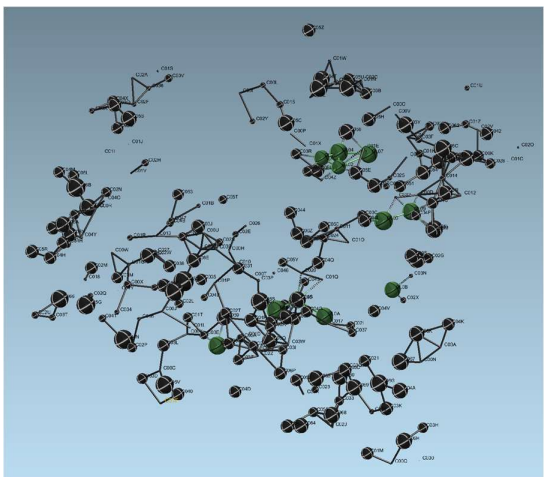

**S8. Currently a proper comparison between our work and Gorelick *et al.*'s work cannot be made**

Though our work has been inspired by Gorelick *et al.*'s work (Gorelick *et al.*, 2023), a proper comparison between these works cannot be made at this point. We use the pR1 method (a model search technique) to directly solve a structure, while Gorelick *et al.* use the molecular-replacement (MR) method as implemented in *phaser* (McCoy *et al.*, 2007), which not only employs a model searching technique (the maximum log-likelihood gain, e.g. LLG), but also employs some phase improvement mechanisms like the dual space cycling, etc. A proper comparison between the pR1 and the LLG can be made if someone tries to use the LLG to directly solve a structure, or, some expert can replace the LLG target by the pR1 target in the MR of the *phaser* program, until then, a proper comparison between our work and Gorelick *et al.*'s work cannot be made.

**References**

Gorelik, T. E., Lukat, P., Kleeberg, C., Blankenfeldt, W. & Mueller, R. (2023). *Acta Cryst*. A**79**, 504-514.

McCoy, A. J., Grosse-Kunstleve, R. W., Adams, P. D., Winn, M. D., Storoni, L. C. & Read, R. J. (2007). *J. Appl. Cryst*. **40**, 658–674.

Zhang, X. & Donahue, J. P. (2024). https://doi.org/10.1107/S2053273324001554.